\documentclass[onecolumn,amsfonts,amssymb,amsmath,floatfix]{article}
\usepackage[dvips]{graphicx}
\usepackage{natbib}
\usepackage{bm}

\title{Scalar potential model of galaxy central mass and central velocity dispersion }

\author{J.C. Hodge$^{1}$\thanks{E-mail:jch9496@blueridge.edu}\thanks{Visiting from XZD Corp., 16 Hosta Ln., Brevard, NC, 28712, E-mail:{scjh@citcom.net}}\\
$^{1}${Blue Ridge Community College, 100 College Dr., Flat Rock, NC, 28731-1690}}

\date{\today}

\begin{document}

\maketitle

\begin{abstract}
The galaxy central mass $M_\mathrm{c}$ and central velocity dispersion $\sigma_\mathrm{c}$ have been found to correlate with large scale galaxy parameters for samples of galaxies with a limited range of characteristics.  A scalar potential model (SPM) that derived from considerations of galaxy clusters, of redshift, of discrete redshift, of H{\scriptsize{I}} rotation curves (RCs) of spiral galaxies and of RC asymmetry is applied to central region parameters.  The $\sigma_\mathrm{c}$ and $ M_\mathrm{c}$ are found to correlate to the host galaxy's and neighboring galaxy's B band luminosity.  The sample included galaxies with rising, flat and declining RCs; galaxies with a wide range of characteristics; and galaxies excluded from samples of other studies of $\sigma_\mathrm{c}$ relationships.  The equations have the same form as the SPM equations for the parameters of the H{\scriptsize{I}} RCs.  Because the SPM is consistent with $M_\mathrm{c}$ and $\sigma_\mathrm{c}$ observations of the sample galaxies, the Sources and Sinks act as monopoles at the center of the galaxies around them.  This suggests the outward scalar potential force of a Source holds the $M_\mathrm{c}$ from collapse into a supermassive black hole.
\end{abstract}

galaxies:fundamental parameters -- galaxies:kinematics and dynamics -- galaxies:nuclei -- cosmology:theory 

\section[SPM of $M_\mathrm{c}$ and $\sigma_\mathrm{c}$]{INTRODUCTION}

Because the amplitude and shape of galaxy rotation curves (RCs) correlate with galaxy luminosity \citep{burn,cati,hodg2,pers}, relationships between galaxy central parameters and large scale galaxy parameters are unexpected by Newtonian dynamics.  

\citet{whit79} and \citet{whit81} found the ratio of the rotation velocity $v_\mathrm{c}$ (km~s$^{-1}$) in the flat region of the RC and the central velocity dispersion $\sigma_\mathrm{c}$ (km~s$^{-1}$) $\approx 1.7$ for a sample of S0 and spiral galaxies.  \citet{gerh} found the maximum circular velocity of giant, round, and nearly non-rotating elliptical galaxies is correlated to the $\sigma _\mathrm{c}$.  \citet{fer4} discovered a power law relationship between circular velocity $v_\mathrm{c25}$ (km~s$^{-1}$) beyond the radius $R_{25}$ of the 25$^{th}$ isophote and $\sigma_\mathrm{c}$ for a sample that also include elliptical galaxies (see her Fig. 1).  \citet{baes} expanded on the data for spiral galaxies with flat and smooth RCs.  {NGC 0598} was a clear outlier.  The $v_\mathrm{c25}$ for {NGC 0598} used in \citet{fer4} was 135 km$\,$s$^{-1}$ that is the highest data point of a rising RC~\citep{corb}.  Galaxies with $\sigma_\mathrm{c} < 70$ km~s$^{-1}$ ($v_\mathrm{c25} < 150$ km$\,$s$^{-1}$) also deviate from the linear relation.  {NGC 4565} was excluded because of warps in the H{\scriptsize{I}} disk.  Also, galaxies with significant non-circular motion of the H{\scriptsize{I}} gas such as {NGC 3031}, {NGC 3079}, and {NGC 4736} were omitted in \citet{fer4}.  {NGC 3200} and {NGC 7171} were also excluded from \citet{fer4} because of discrepant $\sigma_\mathrm{c}$ values in the literature.  

\citet{pizz05} found results similar to \citet{fer4} for high surface brightness (HSB) galaxies.  \citet{pizz05} also found the data consistent with a linear $v_\mathrm{c}$--$\sigma_\mathrm{c}$ relation.  HSB galaxies with flat RCs were chosen for the sample.  Also, galaxies with highly asymmetric RCs and galaxies with RCs not characterized by an outer flat portion were excluded.  Also, \citet{pizz05} found the $v_\mathrm{c}$--$\sigma_\mathrm{c}$ linear relation for low surface brightness (LSB) galaxies is offset with a larger $v_\mathrm{c}$ intercept relative to HSB galaxies.  \citet{buyl} confirmed this distinction between HSB and LSB galaxies for \mbox{$\sigma_\mathrm{c}>80$ km s$^{-1}$} and found that the distinction is less pronounced for galaxies with \mbox{$\sigma_\mathrm{c}<80$ km s$^{-1}$}.  They concluded that the scatter of the $v_\mathrm{c}$--$\sigma_\mathrm{c}$ relation is a function of galaxy mass or that the $v_\mathrm{c}$--$\sigma_\mathrm{c}$ relation changes at \mbox{$\sigma_\mathrm{c} \approx 80$ km s$^{-1}$}.  \citet{hodg2} suggested the offset of the measurement of $v_\mathrm{c}$ between HSB and LSB galaxies is because the measurement is in an area of the rotation curve (RC) more influenced by neighboring galaxies and in an area of the generally rising RC of LSB galaxies.  Further, the criteria for excluding galaxies describe characteristics of large influence of neighbor galaxies.  Using the $v_\mathrm{rrmax}$ as defined by \citet{hodg2} may be more appropriate. 

The masses of compact stellar clusters at the center of low- and intermediate-luminosity galaxies also correlate with the mass of the host galaxy \citep{fer7,wehn}.  \citet{fer7} suggested the compact stellar clusters and the supermassive black hole (SBH) modeled as being at the center of high-luminosity galaxies should be grouped together under the terminology of ``Central Massive Objects'' (CMOs) with mass $M_\mathrm{cmo}$.  The finding of the correlation between $M_\mathrm{cmo}$ and the total mass in a galaxy $M_\mathrm{gal}$ suggests a similar galaxy formation process \citep{fer7,wehn}.

\citet{ghez} and \citet{fer5} have observed Keplerian motion to within one part in 100 in elliptical orbits of stars that are from less than a pc to a few 1000 pc from the center of galaxies.  The stars within nine light hours of the Galaxy center have velocities of 1300 km$\,$s$^{-1}$ to 9000 km$\,$s$^{-1}$\citep{scho} and high accelerations~\citep{ghez}.  A huge amount of mass $M_\mathrm{c}$ (M$_\odot$) such as millions of black holes, dense quark stars~\cite[and references therein]{pras}, and ionized iron \citep{wang} must be inside the innermost orbit of luminous matter \citep{dunn,ghez,ghez3,ghez4,scho}.  

The $M_\mathrm{c}$ varies among galaxies from $10^6 $ M$_\mathrm{\odot}$ to $10^{10}$ M$_\mathrm{\odot}$ \citep{fer2,gebh}.  \citet{fer4} found the ratio of the $M_\mathrm{c}$ to the mass $M_\mathrm{DM}$ of the dark matter halo thought to be around spiral galaxies is a positive value that decreased with $M_\mathrm{DM}$.  The $M_\mathrm{c}$ can be distributed over the central volume with a density of at least $10^{12}$ M$_\mathrm{\odot}$ pc$^{-3}$~\citep{dunn,ghez2,ghez,ghez4}.  The orbits of stars closest to the center of the Galaxy are approximately 1,169 times the Schwartschild radius of a supermassive black hole (SBH) thought to be at the center of the Galaxy~\citep{ghez,scho}.  The orbits of stars closest to the center of the Galaxy are following elliptical paths \citep{ghez} that suggests a net, attractive central force consistent with the Newtonian spherical property \citep{ghez3,scho03}.

That $M_\mathrm{c}$ is crowded into a ball with a radius of less than 45 AU is proven \citep{ghez4}.  That the structure of $M_\mathrm{c}$ is a SBH is widely accepted, but unproven (see \citealt{korm} for a discussion).  The Newtonian model implies the $M_\mathrm{c}$ must either quickly dissipate or must quickly collapse into a SBH~\citep{korm, mago}.  The long term maintenance of $M_\mathrm{c}$ rules out the first possibility.  \citet{moua} suggested there is some extended mass around Sgr A.  Observations have ruled out many models of the nature of $M_\mathrm{c}$ of galaxies \citep{ghez3,scho03}.  

Observations inconsistent with the SBH model include shells of outward flowing, shocked gas around galactic nuclei~\cite[page 595]{binn}\citep{koni}.  \citet{shu} and \citet{silk} suggested a repulsive force, called a ``wind'' (a gas), exerted a repulsive force acting on the cross sectional area of particles.  Therefore, denser particles such as black holes move inward relative to less dense particles.  Less dense particles such as hydrogen gas move outward.  Other observations inconsistent with the SBH model include the apparent inactivity of the central SBH~\citep{baga,baga2,naya,zhao} and the multitude of X-ray point sources, highly ionized iron, and radio flares without accompanying large variation at longer wavelengths reported near the center of the Milky Way \citep{baga,baga2,baga3,binn,genz,zhao,wang}. 

The $M_\mathrm{c}$ correlation with Blue band luminosity $L_\mathrm{bulge}$ of the host galaxy's bulge \citep{korm} has a large scatter.  The $M_\mathrm{c} \propto \sigma_\mathrm{c}^\alpha$, where $\alpha$ varies between 5.27$\pm$0.40~\citep{fer2} and 3.75$\pm$0.3~\citep{gebh}.  The $M_\mathrm{c} - \sigma_\mathrm{c}$ relation appears to hold for galaxies of differing Hubble types, for galaxies in varying environments, and for galaxies with smooth or disturbed morphologies.  \citet[and references therein]{trem} suggested the range of $\alpha$ is caused by systematic differences in the velocity dispersions used by different groups.  \citet{merr4} found the range of $\alpha$ is partly due to the type of regression algorithm used and partly due to the velocity dispersion of the Galaxy sample selected.  Also, discrepancies have been noted among the methods used to measure $M_\mathrm{c}$~\cite[and references therein]{gebh2,merr2}.  \citet{bern} found a selection bias or large scatter in the $M_\mathrm{c}$--$\sigma$ and $M_\mathrm{c}$--$L_\mathrm{bulge}$ correlations that may be the result of more fundamental relations among $M_\mathrm{c}$, $\sigma$, and $L_\mathrm{bulge}$.

A scalar potential model (SPM) was derived from considerations of galaxy clusters \citep{hodg}.  The SPM suggests the RCs of spiral galaxies are determined by a scalar potential term added to the conventional Newtonian rotation velocity equation.  The scalar potential term is proportional to Blue band luminosity $L$ (erg s$^{-1}$) of a galaxy.  For spiral galaxies (Sources), the scalar potential term is directed outward.  For other galaxies (Sinks), the scalar potential term is directed inward.  The SPM found parameters $P$ of H{\scriptsize{I}} RCs of spiral galaxies are related to $L$ of the host galaxy and of nearby galaxies \citep{hodg2}.  The parameters are the square of the rotation velocity, the radius, the mass, and the acceleration at discontinuities in the RC.  The equation is
\begin{eqnarray}
\frac{P}{unit} = & K_\mathrm{1} \, B_\mathrm{1}^{I_1} \, \frac{L}{10^8\, \mathrm{erg}\,\mathrm{s}^{-1}} + \nonumber \\*
 &(-1)^{s} K_\mathrm{2} B_\mathrm{2}^{I_2} \frac{\vert \vec{K} \bullet \vec{a}_\mathrm{o} \vert} {10^3 \,  \mathrm{kpc}^{-1} \, \mathrm{km}^2 \, \mathrm{s}^{-2}} \pm \sigma_\mathrm{e} 
\label{eq:1},
\end{eqnarray}
where $unit$ is the units of $P$; $ K_{ 1}$, $ K_{ 2}$, $B_\mathrm{1}$, and $ B_\mathrm{2}$ are constants that are unique for each $P$; $I_1$ and $I_2$ are integers that are unique for each galaxy; $\vert \vec{K} \bullet \vec{a}_\mathrm{o} \vert $ is the influence of nearby galaxies and is a correction term to the primary $P - L$ relationship; $s$ determines the sign of the $\vert \vec{K} \bullet \vec{a}_\mathrm{o} \vert $ term; $\vec{K}$ is a constant vector common for all galaxies; $\vec{a}_\mathrm{o}$ is the acceleration vector that is calculated from the orientation of the host galaxy, the $L$ of the neighboring galaxies, and the relative position of the neighboring galaxies; and $\sigma_\mathrm{e}$ is the standard deviation of the relative differences ($\delta P/P$) of the sample galaxies.  

This Paper pursues the possibility of a relation between $M_\mathrm{c}$, $\sigma_\mathrm{c}$, and $L$ of the host and neighboring galaxies suggested by the SPM.  A correlation is found in the form of Eq.~(\ref{eq:1}).  Therefore, a central, repulsive force $F_\mathrm{s}$ exerted by the scalar potential $\rho$ field exists to maintain the $M_\mathrm{c}$ of spiral galaxies from collapse.  

In section~\ref{sec:sample}, the sample is described.  Equation~(\ref{eq:1}) is used to calculate $M_\mathrm{c}$ and $\sigma_\mathrm{c}$ in Section~\ref{sec:results}.  The discussion and conclusion are in Section~\ref{sec:disc}.  

\section[SPM of $M_\mathrm{c}$ and $\sigma_\mathrm{c}$]{\label{sec:sample}Sample}

The galaxies used in the calculations were those used in \citet{hodg2}.  That is, they were selected from the NED database\footnote{The Ned database is available at http://nedwww.ipac.caltech.edu.  The data were obtained from NED on 5 May 2004.}.  The selection criteria were that the heliocentric redshift $z_\mathrm{h}$ be less than 0.03 and that the object be classified as a galaxy.  The parameters obtained from the NED database included the name, equatorial longitude $E_\mathrm{lon}$ (degrees) for J2000.0, equatorial latitude $E_\mathrm{lat}$ (degrees) for J2000.0, morphology, the B-band apparent magnitude $m_\mathrm{b}$ (mag.), and the extinction $E_{\mathrm{xt}}$ (mag.) as defined by NED.  The galactocentric redshift $z$ was calculated from $z_\mathrm{h}$. 

The $\sigma_\mathrm{c}$, the 21-cm line width $W_\mathrm{20}$ (km~s$^{-1}$) at 20 percent of the peak, the inclination $i_\mathrm{n}$ (arcdegrees), and the position angle $p_\mathrm{a}$ (arcdegrees) for galaxies were obtained from the LEDA database\footnote{The LEDA database is available at http://leda.univ-lyon.fr.  The data were obtained from LEDA on 5 May 2004.} if such data existed. 

The host sample galaxies with $\sigma_\mathrm{c}$, $m_\mathrm{b}$, $W_\mathrm{20}$, $i_\mathrm{n}$, and $p_\mathrm{a}$ values were (1) those used in \citet{hodg2} from \citet{bege,broe,free,garc,guha,korn,korn2,lisz,macr,mann,rubi85}; and \citet{swat}, (2) those used in \citet{fer4}, and (3) those specifically excluded from \citet{fer4}.  A total of 82 host sample galaxies were used for the $\sigma_\mathrm{c}$ calculation.  Of the host galaxies, 60 are Source galaxies and 22 are Sink galaxies.  Tables \ref{tab:1} and \ref{tab:2} lists the host galaxies used in the $\sigma_\mathrm{c}$ calculation.  Table \ref{tab:2} lists the 29 host galaxies used in the $M_\mathrm{c}$ calculation.  

\begin{table}
\tiny
\caption{Data for the host sample galaxies used in the $\sigma_\mathrm{c}$ calculation. }
\label{tab:1} 
\begin{tabular}{llrrrrrr}
\hline
Galaxy&Morphology&$L$$^\mathrm{a}$&$\vert K \bullet a \vert$$^\mathrm{b}$&$\sigma_\mathrm{c}$$^\mathrm{c}$&$m_1$&$m_2$&$\delta \sigma_\mathrm{c} / \sigma_\mathrm{c}$\\
\hline
{IC 0342}&SAB(rs)cd HII&4.014&$^\mathrm{d}$&74&0&$^\mathrm{d}$&-0.11\\
{IC 0724}&Sa &1.796&0.057&246&5&11&-0.02\\
{N 0224}&SA(s)b LINER&1.125&20.414&170&5&6&0.03\\
{N 0598}&SA(s)cd HII&0.219&27.936&37&2&1&0.01\\
{N 0701}&SB(rs)c Sbrst &0.425&1.198&73&3&6&0.02\\
{N 0753}&SAB(rs)bc &1.255&0.584&116&3&6&0.00\\
{N 0801}&Sc &1.162&0.753&146&4&7&0.00\\
{N 1024}&(R')SA(r)ab &1.714&0.713&173&4&8&-0.02\\
{N 1353}&SA(rs)bc LINER &1.014&0.386&87&3&9&0.17\\
{N 1357}&SA(s)ab &2.063&0.270&124&3&10&0.08\\
{N 1417}&SAB(rs)b &1.580&0.613&140&3&8&0.05\\
{N 1515}&SAB(s)bc &0.745&1.896&101&4&7&-0.11\\
{N 1620}&SAB(rs)bc &1.120&0.446&124&4&9&-0.11\\
{N 2639}&(R)SA(r)a ? Sy1.9 &4.588&0.023&198&3&13&0.04\\
{N 2742}&SA(s)c &0.750&2.353&66&2&4&-0.01\\
{N 2775}&SA(r)ab &2.987&0.103&176&3&7&0.00\\
{N 2815}&(R')SB(r)b &2.217&3.306&203&4&6&0.01\\
{N 2841}&SA(r)b ;LINER Sy1 &1.822&0.524&206&4&9&0.04\\
{N 2844}&SA(r)a &0.602&0.055&110&4&9&-0.01\\
{N 2903}&SB(s)d HII &0.947&0.130&102&3&8&0.02\\
{N 2998}&SAB(rs)c &1.022&2.551&91&3&6&-0.06\\
{N 3067}&SAB(s)ab? HII &0.309&0.002&80&4&12&-0.01\\
{N 3079}&SB(s)c;LINER Sy2 &1.300&0.036&146&4&11&-0.07\\
{N 3145}&SB(rs)bc &1.529&0.452&169&4&8&0.03\\
{N 3198}&SB(rs)c &0.855&0.272&63&2&7&-0.08\\
{N 3200}&SA(rs)bc &1.930&0.009&177&4&14&0.12\\
{N 3593}&SA(s)0/a;HII Sy2 &0.263&0.059&54&3&8&0.01\\
{N 4051}&SAB(rs)bc Sy1.5 &2.166&0.653&84&1&7&-0.04\\
{N 4062}&SA(s)c HII &0.518&0.869&93&4&7&0.00\\
{N 4216}&SAB(s)b HI I/LINER &1.731&0.175&207&4&11&-0.05\\
{N 4321}&SAB(s)bc;LINER HII &2.209&2.613&86&1&5&0.04\\
{N 4378}&(R)SA(s)a Sy2 &6.049&0.120&198&2&11&-0.04\\
{N 4388}&SA(s)b sp Sy2 &0.791&4.067&115&4&6&0.04\\
{N 4414}&SA(rs)c? LINER &1.294&2.158&110&3&6&0.01\\
{N 4448}&SB(r)ab &1.163&1.000&173&5&9&-0.07\\
{N 4548}&SBb(rs);LINER Sy &1.087&0.002&144&4&12&0.00\\
{N 4565}&SA(s)b? sp Sy 3 Sy1.9 &1.655&12.922&136&3&4&0.01\\
{N 4647}&SAB(rs)c &0.748&0.134&98&3&9&0.05\\
{N 4698}&SA(s)ab Sy2 &1.396&3.587&133&3&6&0.05\\
{N 4736}&(R)SA(r)ab;Sy 2 LINER &1.230&1.092&104&3&7&-0.02\\
{N 4866}&SA(r)0+ sp LINER &1.904&0.416&210&4&10&-0.08\\
{N 5033}&SA(s)c Sy1.9 &1.302&0.937&131&3&8&0.00\\
{N 5055}&SA(rs)bc HI I/LINER &1.383&3.582&101&3&7&0.21\\
{N 5297}&SAB(s)c sp &0.956&1.432&119&4&8&0.11\\
{N 5457}&SAB(rs)cd &2.129&0.370&73&1&7&-0.05\\
{N 6503}&SA(s)cd HI I/LINER &0.197&0.303&46&3&5&-0.01\\
{N 6814}&SAB(rs)bc Sy1.5&0.036&2.830&112&9&7&0.05\\
{N 7171}&SB(rs)b &1.006&0.457&84&2&8&-0.04\\
{N 7217}&(R)SA(r)ab;Sy LINER &2.117&0.918&127&3&8&-0.11\\
{N 7331}&SA(s)b LINER &1.570&0.711&138&3&8&-0.01\\
{N 7506}&(R')SB(r)0+ &0.741&0.051&147&5&12&0.15\\
{N 7537}&SAbc &0.693&146.761&78&3&1&-0.01\\
{N 7541}&SB(rs)bc pec HII &1.397&125.361&65&1&-1&0.01\\
\hline
\end{tabular}

$^\mathrm{a}$ Unit: $10^8\, \mathrm{erg\,cm}^{-2}\,\mathrm{s}^{-1}$.\\
$^\mathrm{b}$ Unit: $10^3 \,  \mathrm{kpc}^{-1} \, \mathrm{km}^2 \, \mathrm{s}^{-2}$.\\
$^\mathrm{c}$ Unit: $10^3$ km$^2$ s$^{-2}$.\\
$^\mathrm{d}$ This galaxy has a $z$ value too small to obtain the $\vert K \bullet a \vert$ measurement.
\end{table}

\begin{table}
\tiny
\caption{Data for the host sample galaxies used in the $\sigma_\mathrm{c}$ and $M_\mathrm{c}$ calculations. }
\label{tab:2} 
\begin{tabular}{llrrrrrrrrrr}
\hline
Galaxy&Morphology&$L$$^\mathrm{a}$&$\vert K \bullet a \vert$$^\mathrm{b}$&$\sigma_\mathrm{c}$$^\mathrm{c}$&$M_\mathrm{c}$$^\mathrm{d}$&$m_1$&$m_2$&$\delta \sigma_\mathrm{c} / \sigma_\mathrm{c}$&$n_1$&$n_2$&$\delta M_\mathrm{c} / M_\mathrm{c}$\\
\hline
{I 1459}&E3 LINER &3.757&0.809&306&4.600&4&10&0.00&5&10&-0.02\\
{N 0221}&cE2 &0.021&$^\mathrm{e}$&72&0.039&8&&0.03&6&$^\mathrm{e}$&0.05\\
{N 2787}&SB(r)0+ LINER &0.122&2.230&194&0.410&8&8&-0.04&7&2&-0.01\\
{N 3031}&SA(s)ab;LINER Sy1.8&1.047&0.352&162&0.680&4&10&-0.11&4&6&0.01\\
{N 3115}&S0- &1.187&322.121&252&9.200&6&11&0.05&9&13&-0.03\\
{N 3245}&SA(r)0\^{ }0\^{ } ?;H IILINER &0.956&6.161&210&2.100&5&7&-0.02&6&5&-0.04\\
{N 3379}&E1 LINER &0.979&14.051&207&1.350&5&6&-0.05&6&3&-0.01\\
{N 3608}&E2 LINER &1.164&0.708&192&1.100&5&9&0.05&5&6&-0.01\\
{N 4258}&SAB(s)bc;LINE R Sy1.9 &1.188&1.536&134&0.390&4&8&0.07&3&2&0.02\\
{N 4261}&E2-3 LINER &2.972&10.588&309&5.400&5&7&0.04&6&1&0.00\\
{N 4342}&S0- &0.121&0.108&251&3.300&9&11&0.01&11&5&-0.01\\
{N 4374}&E1;LERG LINER &3.595&9.080&282&17.000&4&7&-0.06&8&7&-0.02\\
{N 4473}&E5 &0.941&0.842&179&0.800&5&8&0.04&5&7&0.01\\
{N 4486}&E+0-1 pec;NLR g Sy &5.075&0.084&333&35.700&4&12&-0.02&8&19&0.00\\
{N 4564}&E6 &0.362&56.704&157&0.570&6&0&0.00&6&-3&-0.01\\
{N 4649}&E2 &3.699&1.337&335&20.600&5&9&-0.07&8&9&0.00\\
{N 4697}&E6 &1.291&1.994&174&1.700&4&7&0.07&6&8&0.07\\
{N 5128}&S0 pec Sy2 &0.655&1.739&120&2.400&4&6&0.04&7&7&0.00\\
{N 5845}&E &0.368&0.314&234&2.900&7&10&-0.02&9&11&0.01\\
{N 6251}&E;LERG Sy2&4.990&0.016&311&5.900&4&10&0.00&5&17&0.03\\
{N 7052}&E&3.284&0.232&270&3.700&4&10&0.04&5&11&-0.02\\
{N 3384}&SB(s)0-&0.690&2.424&148&0.140&5&7&0.03&2&-4&0.00\\
{N 4742}&E4 &0.193&7.798&109&0.140&6&5&0.04&4&-1&-0.02\\
{N 1023}&SB(rs)0- &0.625&2.266&204&0.440&6&6&0.00&4&3&0.01\\
{N 4291}&E3 &0.816&1.414&285&1.900&7&9&-0.11&7&8&-0.06\\
{N 7457}&SA(rs)0-? &0.323&0.900&69&0.036&4&7&0.00&1&-2&0.02\\
{N 0821}&E6? &1.518&0.329&200&0.390&5&10&-0.15&3&8&-0.04\\
{N 3377}&E5-6 &0.454&0.442&139&1.100&5&8&0.04&7&9&-0.03\\
{N 2778}&E &0.253&2.230&162&0.130&7&7&-0.10&4&1&0.01\\
\hline
\end{tabular}

$^\mathrm{a}$ Unit: $10^8 \, \mathrm{erg\,cm}^{-2}\,\mathrm{s}^{-1}$.\\
$^\mathrm{b}$ Unit: $10^3 \,  \mathrm{kpc}^{-1} \, \mathrm{km}^2 \, \mathrm{s}^{-2}$.\\
$^\mathrm{c}$ Unit: $10^3$ km$^2$ s$^{-2}$.\\
$^\mathrm{d}$ Unit: $10^8 \, M_\odot$.\\
$^\mathrm{e}$ This galaxy has a $z$ value too small to obtain the $\vert K \bullet a \vert$ measurement. 
\end{table}

The distance $D$ (Mpc) data for the 29 host sample used in the $M_\mathrm{c}$ calculation were taken from \citet{merr2}.  The $D$ to nine host galaxies was calculated using Cepheid stars from \citet{free} and \citet{macr}.  The $D$ to {NGC 3031} and {NGC 4258} were from \citet{merr2} rather than from \citet{free}.  The $D$ to the remaining host sample galaxies was calculated using the Tully-Fisher relation with the constants developed in \citet{hodg}.  The remaining galaxies from the NED database were neighbor galaxies.  The $D$ of these galaxies was calculated from the relative $z$ and $D$ of the host galaxy as described by \citet{hodg2}.  The $L$ for the galaxies was calculated from $D$, $m_\mathrm{b}$, and $E_\mathrm{xt}$. 

This host galaxy sample has LSB, medium surface brightness (MSB), and HSB galaxies; includes LINER, Sy, HII, and less active galaxies; field and cluster galaxies; galaxies with rising, flat, and declining RCs; and galaxies with varying degrees of asymmetry.  The host sample includes {NGC 0598}, {NGC 3031}, {NGC 3079}, {NGC 3200}, {NGC 4565}, {NGC 4736}, and {NGC 7171} that were excluded from \citet{fer4}, six galaxies with $\sigma_\mathrm{c} < 70$ km~s$^{-1}$, and galaxies that \citet{pizz05} would exclude.  

\section[SPM of $M_\mathrm{c}$ and $\sigma_\mathrm{c}$]{\label{sec:results}Results}

Appling the same procedure used by \citet{hodg2} for finding the parametric equations yields: 
\begin{eqnarray}
\frac{\sigma_\mathrm{c}^2}{10^3 \, \mathrm{km}^2 \, \mathrm{s}^{-2}} = & K_\mathrm{\sigma 1} \, B_\mathrm{\sigma 1}^{m_1} \, \frac{L}{10^8\, \mathrm{erg}\,\mathrm{s}^{-1}} + \nonumber \\*
 &(-1)^{s} K_\mathrm{\sigma 2} B_\mathrm{\sigma 2}^{m_2} \frac{\vert \vec{K} \bullet \vec{a}_\mathrm{o} \vert} {10^3 \,  \mathrm{kpc}^{-1} \, \mathrm{km}^2 \, \mathrm{s}^{-2}} \pm 6\% 
\label{eq:2}
\end{eqnarray}
and 
\begin{eqnarray}
\frac{M_\mathrm{c}}{10^8 \, M_\odot} = & K_\mathrm{M1} \, B_\mathrm{M1}^{n_1} \, \frac{L}{10^8\, \mathrm{erg}\,\mathrm{s}^{-1}} + \nonumber \\*
 &(-1)^{s_\mathrm{M}} K_\mathrm{M2} B_\mathrm{M2}^{n_2} \frac{\vert \vec{K} \bullet \vec{a}_\mathrm{o} \vert} {10^3 \,  \mathrm{kpc}^{-1} \, \mathrm{km}^2 \, \mathrm{s}^{-2}} \pm 3\% 
\label{eq:3},
\end{eqnarray}
where $ K_{\sigma 1}=1.6 \pm 0.3$, $ K_{\sigma 2}= (2.5 \pm 0.7) \times 10^{-3}$, $B_\mathrm{\sigma 1}= 1.88 \pm 0.06$, $ B_\mathrm{\sigma 2} = 2.52 \pm 0.09$, $ K_{M 1} =0.7 \pm 0.1$, $ K_{M 2} = (5.1 \pm 0.8) \times 10^{-3}$, $B_\mathrm{M 1}=1.73 \pm 0.04$, and $ B_\mathrm{M 2} = 1.70 \pm 0.03$. The integer values for each host sample galaxy are listed in Tables~\ref{tab:1} and \ref{tab:2}.  In both relations, $\sigma_\mathrm{e} = 18\%$ for the $L$ term, only.  

Equation~(\ref{eq:1}) was derived assuming the Newtonian spherical property applied.  That is, the Source was internal to the orbits of the mass under consideration.  The applicability of Eq.~(\ref{eq:1}) to $\sigma_\mathrm{c}$ and $ M_\mathrm{c}$ suggests the Source or Sink acts as a monopole internal to the particles of $M_\mathrm{c}$.

\section[SPM of $M_\mathrm{c}$ and $\sigma_\mathrm{c}$]{\label{sec:disc}Discussion and conclusion}

The SPM speculates structures of the central mass and the structure of stellar nuclear clusters are the same.  The suggested CMO structure is a central Source of a matter-repulsive $\rho \propto R^{-1}$, where $R$ is the galactocentric radius, surrounded by a spherical shell of matter.  The SPM suggests the $L \propto \epsilon$, where $\epsilon$ is the Source strength, and, therefore, $F_\mathrm{s} \propto \nabla \rho$ at a given $R$ on the cross section of matter $ m_\mathrm{s}$.  Therefore, the density ($m_\mathrm{s}/m_\mathrm{i}$), where $m_\mathrm{i}$ is the inertial mass, of particles at a given radius varies with $L$.  Therefore, the galaxies with larger $L$ will have more mass in the center shell to balance the higher $F_\mathrm{s}$ with the gravitational force $F_\mathrm{g}$.  Therefore, the SPM naturally leads to the smoothness of the $M_\mathrm{cmo}$ -- $M_\mathrm{gal}$ relation for the full range of CMO spiral galaxies.

If this speculation is essentially correct, then the correlation of central parameters with spiral galaxy global and RC parameters suggests not only a similar galaxy formation process but also a self- regulatory, negative feedback process continually occurring.  Feedback processes have been suggested in several recent studies of galaxies with CMOs (e.g. \citealt{li06,merr2,robe}).  I further speculate the $\epsilon$ is the control of the negative feedback process.  If the mass of the CMO increases, the $F_\mathrm{g}$ increases and mass migrates inward.  At very high $\rho$, the high repulsive $F_\mathrm{s}$ compresses matter, the mass (black hole) cracks like complex molecules in the high heat and pressure of a fractional distillation process, and matter is reclaimed as radiation and elementary particles that form hydrogen.  This accounts for the large amount of hydrogen outflowing from the Galaxy center and shocked gas near the Galaxy center.  A single black hole reclamation event is consistent with the periodic X-ray pulses from the Galaxy center.  Further, the feedback loop controlled by $\epsilon$ is the connection among the central parameters, outer RC parameters, and the global parameters of spiral galaxies.  However, the $\epsilon$ of a galaxy acts only radially.  Therefore, the $\vert \vec{K} \bullet \vec{a}_\mathrm{o} \vert $ terms effects are the asymmetry and the formation, evolution, and maintenance of the rotation of particles.  This effect may be calculated only if the classification of parameters is first calculated.

Another speculation is that there may be galaxies with higher and lower values of $\epsilon$ than in spiral galaxies.  For instance, QSOs may have a higher value of $\epsilon$ that ejects matter from a spiral configuration (e.g see the images of \citealt{sule}).  A smaller value of $\epsilon$ would be insufficient to form a disk. 

The $L$ term is the primary, determining factor of the parameter relations.  The neighboring galaxies cause the scatter noted in previous studies.  The special focus of the present investigation included galaxies that are problematic in other models.  Considering the range of observations and range of galaxy characteristics with which the SPM is consistent, the SPM is a relatively simple model. 

The SPM was applied to central region parameters.  For a sample of 60 Source galaxies and 22 Sink galaxies, the $\sigma_\mathrm{c}$ was found to correlate to the host galaxy's and neighboring galaxy's B band luminosity.  The sample included galaxies with rising, flat and declining RCs; galaxies with a wide range of characteristics; and galaxies excluded from samples of other studies of $\sigma_\mathrm{c}$ relationships.  For a sample of seven Source galaxies and 22 Sink galaxies, the $ M_\mathrm{c}$ was found to correlate to the host galaxy's and neighboring galaxy's B band luminosity.  The equations have the same form as the SPM equations for the parameters of the H{\scriptsize{I}} RCs.  The Sources and Sinks act as monopoles at the center of the galaxies around them.  The SPM is consistent with $M_\mathrm{c}$ and $\sigma_\mathrm{c}$ observations of the sample galaxies.

\section*{Acknowledgments}
This research has made use of the NASA/IPAC Extragalactic Database (NED) which is operated by the Jet Propulsion Laboratory, California Institute of Technology, under contract with the National Aeronautics and Space Administration.

This research has made use of the LEDA database (http://leda.univ-lyon1.fr).

I acknowledge and appreciate the financial support of Maynard Clark, Apollo Beach, Florida, while I was working on this project.





\end{document}